\documentclass[letterpaper]{article} %DO NOT CHANGE THIS
\usepackage{aaai18}  %Required
\usepackage{times}  %Required
\usepackage{helvet}  %Required
\usepackage{courier}  %Required
\usepackage{url}  %Required
\usepackage{graphicx}  %Required
\usepackage{appendix}
\usepackage{hyperref}

\begin{document}
\title{Applying Machine Learning Methods to Enhance the Distribution of \\
Social Services in Mexico}

\author{Kris Sankaran\textsuperscript{1},
Diego Garcia-Olano\textsuperscript{2},
Mobin Javed\textsuperscript{3},
Maria Fernanda Alcala-Durand\textsuperscript{4}\\
\Large\bf Adolfo De Un\'{a}nue\textsuperscript{4},
Paul van der Boor\textsuperscript{5},
Eric Potash\textsuperscript{5}\\
\Large\bf Roberto S\'{a}nchez Avalos\textsuperscript{6},
Luis I\~{n}aki Alberro Encinas\textsuperscript{6},
Rayid Ghani\textsuperscript{5} \\
\textsuperscript{1}Stanford University,
\textsuperscript{2}University of Texas at Austin,
\textsuperscript{3}UC Berkeley,
\textsuperscript{4}Instituto Tecnol\'{o}gico Aut\'onomo de M\'{e}xico \\
\textsuperscript{5}University of Chicago,
\textsuperscript{6}Secretar\'{i}a de Desarrollo Social\\
kriss1@stanford.edu,
\{diegoolano, mfalcalad, pvboor\}@gmail.com,
mobin.javed@berkeley.edu  \\
\{adolfo, epotash, rayid\}@uchicago.edu
\{roberto.sancheza, luis.alberro\}@sedesol.gob.mx}

\maketitle
\begin{abstract}
The Government of Mexico's social development agency, SEDESOL, is responsible for the administration of social services and has the mission of lifting Mexican families out of poverty. One key challenge they face is matching people who have social service needs with the services SEDESOL can provide accurately and efficiently. In this work we describe two specific applications implemented in collaboration
with SEDESOL to enhance their distribution of social services. The first problem relates to systematic underreporting on applications for social services, which makes it difficult to identify where to prioritize outreach. Responding that five people reside in a home when only three do is a type of underreporting that could occur while a social worker conducts a home survey with a family to determine their eligibility for services. The second involves approximating multidimensional poverty profiles across households. That is, can we characterize different types of vulnerabilities -- for example, food insecurity and lack of health services -- faced by those in poverty?

We detail the problem context, available data, our machine learning formulation, experimental results, and effective feature sets. As far as we are aware this is the first time government data of this scale has been used to combat poverty within Mexico. We found that survey data alone can suggest potential underreporting.

Further, we found geographic features useful for housing and service related indicators and transactional data informative for other dimensions of poverty.
The results from our machine learning system for estimating poverty profiles will directly help better match 7.4 million individuals to social programs.
\end{abstract}

\section{Introduction}

Secreter\'{i}a de Desarrollo Social (SEDESOL) is the Government of Mexico's social development agency and is responsible for providing social services in Mexico and reducing poverty in the country. A key challenge they face is matching people who have social service needs with the services SEDESOL can provide accurately and efficiently. This paper is focused on a collaboration with SEDESOL focusing on helping them tackle two related problems that would help them more effectively match social services with those most in need
by 1) identifying individuals who are submitting potentially false information in order to get services they're eligible for and 2) identifying individuals who have unmet needs that SEDESOL can meet.

\begin{enumerate}

\item The first problem is related to systematic underreporting on applications for social services --  flagging suspicious applications and facilitate investigation of those providing false responses to qualify for services including conditional cash-transfer programs. These investigations can help  ensure that these services, which sometimes have limited resources and long waiting lists, go to those in real need.

\item The second problem, also related to improving the distribution of services, estimate the multidimensional poverty profile 
of a household -- usually computed using detailed surveys which are time and labor intensive to collect -- using both internal transactional and publicly available 
census data. This information is used to discover individuals with unmet needs and
then direct the relevant service providers to those individuals.
\end{enumerate}

We begin with a brief review of context, describe our general machine learning formulation, and conclude with results and next steps.
We propose different feature sets and evaluation strategies for each problem and explain contrasting situations where our machine learning pipeline
either succeeds or fails in delivering meaningful improvements over natural baselines.
We have made all code publicly available at \href{https://github.com/dssg/sedesol-public}{https://github.com/dssg/sedeol-public}.

\subsection{SEDESOL: The Government of Mexico's social development agency}

SEDESOL operates to fight poverty in Mexico \cite{sedesol_2013}. They aim to break the poverty cycle for communities and individuals by empowering them through nutrition, education, social security, training and employment programs.

The organization in charge of quantifying poverty and evaluating social policy in Mexico is the Consejo Nacional de Evaluaci\'on de la Pol\'tica de Desarrollo Social (CONEVAL). They have developed a multidimensional measure of poverty based on Welfare Income Lines and six Basic Needs Indicators \cite{conevalindicators},

\begin{enumerate}
\item Education: education level or ability to attend school.
\item Health Services: lacking health services, via work benefit, government program or voluntary enrollment.
\item Social Security: lacking access to social security.
\item Quality of the Dwelling: structure and size of home.
\item Basic Services: water, drainage, electricity.
\item Food: food insecurity of any degree.
\end{enumerate}

The Welfare Income Lines are determined first by location (urban or rural) and then by level of welfare. The Minimum Welfare Line (LBM, in Spanish: L\'inea de Bienestar M\'inimo) represents the income level a person needs to cover basic food needs, while the Welfare Line (LB, in Spanish: L\'inea de Bienestar) includes additional expenses related to basic goods and services.

To receive access to programs, potential beneficiaries  take a survey called the Single Questionnaire of Socioeconomical Indicators (CUIS, in Spanish: Cuestionario \'Unico de Indicadores Socioecon\'omicos). The survey contains information about the household, including household characteristics, services and household supplies, and describes individuals inhabiting the household, including education, work situation, and income.

\subsection{Problem description}

Survey responses are used to estimate household income, which must be below the LBM in order to be eligible for receiving assistance. This creates an incentive for underreporting the household situation and supplies, as well as their self-reported income and expenses (these variables are in fact not taken into account while estimating the eligibility conditions) \cite{martinelli2009deception}. Responding that five people reside in a home when only three do is a type of underreporting that could occur while a social worker conducts a home survey with a family to determine their eligibility for services. However, for a subset of the beneficiaries, there is a second part of the survey, called the Home Verification Module, which consists of a surveyor going inside the house of a potential beneficiary who has just taken the survey and verifying the self-reported answers concerning observable housing variables.

In an effort to avoid program redundancy and centralize transactional information, SEDESOL collects all the assistance information from a wide range of social service programs, into the Single Register of Beneficiaries (PUB, in Spanish: Padr\'on \'Unico de Beneficiarios) \cite{gomez2011padron}. However, not all beneficiaries in PUB have taken the CUIS, so the only available information about them is which programs they receive assistance from and where they live.

Using responses from CUIS, there is enough information to determine the presence or absence of each of the six dimensions of poverty. This information
helps SEDESOL better understand the needs of the population it serves. However, since most beneficiaries do not take a CUIS, a natural question is whether the six indicators can be estimated directly from PUB and spatial context.  We call this problem the \textbf{imputation of poverty indicators}.

\section{Existing solutions and related work}

Several systems are in place to address both underreporting and imputation of poverty indicators, though none so far are both reliable
and scalable.

The most direct solution is to require that surveyors attempt to complete a home verification module.
While this provides a validated response for a number of questions, and therefore an indication of whether a respondent underreported, it
can only practically be applied for a small subset of SEDESOL's beneficiaries -- currently 409,000 of the 6.8 million households 
who have taken the CUIS are covered by a home verification module.

An alternative involves comparing self-reported incomes with those estimated by responses from other questions. Since no ground truth is available 
for SEDESOL beneficiaries, this approach relies on surveys developed by CONEVAL to understand regional poverty profiles \cite{conevalindicators}. It is unclear whether those 
surveyed by CONEVAL are representative of the population of SEDESOL beneficiaries, and even if they were, estimating income from limited household characteristics is challenging.

For the imputation problem, a direct solution is increasing the number of individuals who are given comprehensive CUIS. This is not scalable. An approximate solution is provided
by CONEVAL, which estimates multidimensional poverty profiles at the municipal and state levels every five and two years, respectively, based on 
general population surveys \cite{conevalindicators}. For a finer-resolution view of poverty indicators tailored to SEDESOL beneficiaries, further work is needed.

\section{Approach}
Here we describe the data sources used, our problem formulation, features generated, the modeling methods, and an evaluation of the results across those methods. 

\subsection{Data}

We compiled beneficiary enrollment, socioeconomic, and geospatial data for modeling both underreporting and imputation of poverty indicators.

We extracted transactions from PUB associated with the last quarter of 2015. This comprises 120
million payments across households. In addition to beneficiary IDs and payment amounts, the data
specifies the program providing the payment and the home address of the recipient, both potentially associated with household poverty profiles.

In addition to responses for household CUIS surveys\cite{elizaldeinvestigadores}, SEDESOL has estimated six binary variables corresponding to the multidimensional poverty profile. We use these indicators as model responses, and the 
intersection between CUIS and PUB as the basis for imputation modeling. This comprises approximately 18 million individuals, though final training is based only the 7.4 million individuals for which locality information is available. In addition to these surveys, we obtained the home verification module from the subset of programs that require this additional step.

To supplement these program data, we collected related geospatial and socioeconomic information. As 
geospatial data, we generated latitude-longitude coordinates based on the addresses recorded in PUB, using an open source 
geocoding library \cite{ge2005address}. Naive implementations were unsatisfactory, since addresses in Mexico are often not standardized \cite{ackermann2016designing}. 
Best performance was achieved with a composite approach, blending one based on street address searches constrained to known localities, and another using information on known side streets. Evaluation was based on the proportion of geocoded addresses in the correct (a priori known) sublocality.

For socioeconomic context, we retrieved census data from INEGI \cite{inegi2011vivienda}. This data describes
demographics and development across Mexico, at the level of street blocks.

\subsection{Formulation}

There are several ways to frame both the underreporting and imputation problems. Here we describe the reason behind our particular machine learning formulation. 

\subsubsection{Underreporting estimation}
For the underreporting problem, the primary factors
potentially useful as model responses are (1) difference between self-reported and estimated incomes, 
(2) general discrepancies on the home verification module, (3) a binary overall "potentially underreporting" question within the module, also filled out by the surveyor. 
Further, a fourth, unsupervised approach is possible, flagging any outliers in CUIS responses as under-reporters, after having initially matched based on geographic, program, and 
personal characteristics.

The first approach has the advantage of enabling training on the CUIS that are not associated with a home verification visit;
however, estimating income can itself be a difficult problem \cite{bustos2015estimation}. The second and third approaches restrict training to those households
visited by home verification surveyors, but have the advantage of reflecting the validated opinions of trained surveyors, and so were preferred. Between these options,
we choose to use the full set of verification questions and summarize them according to whether there was a discrepancy on any of them.
This method allows us to retrospectively identify which questions are mostly often underreported.
The fourth approach is similar to the first in that it allows training on all CUIS; however, the results would require more nuanced interpretation and follow-up, which would complicate deployment and decrease the chances of the system being used.

Regardless of the response, a household's distance from the poverty line is also relevant when operationalizing results, because this threshold is used to determine program eligibility.
Therefore, even if a household is predicted to be underreporting, if it is significantly below the eligibility threshold, it may not be worth flagging since they're still eligible and will not be investigated further. 
Rather than incorporating this in the model training stage, we account for this fact in the interpretation stage, see Section \ref{sec:evaluation}.

\subsubsection{Imputation of Poverty Indicators}

between features and responses is comparable across both populations. To assess the quality of this approximation would require on-the-ground experiments. Further, while treating responses independently
prevents improvements based on potential correlation between responses, we preferred maintaining access to the much larger class of modeling
techniques designed for single responses, though multitask methods are a potentially interesting area of future investigation.

\subsection{Features}

We use four types of features: spatial, transactional, socioeconomic, and survey features. All are available in the underreporting problem, but survey responses cannot be used for imputation, since they require completion of the CUIS.
In both cases, features are generated at the household level. 

For spatial features, we consider the raw geographic coordinates resulting from geocoding. We also compute averages, restricted to training folds, of each poverty indicator over street blocks and localities. 

For transactional features, we build summaries of a household's PUB transaction history. For example, we consider
which programs they are enrolled in, the number, rate, and total amount of payments coming from these programs, and the initial enrollment date. These are natural choices of features for the 
imputation problem, because certain programs are directed towards specific poverty indicators, and enrollment to these programs often occurs during  program enrollment campaigns.

Socioeconomic features are derived from summaries reported by INEGI at the locality and street-block levels \cite{inegi2011vivienda}. For example, these data 
include the estimated proportion of households within a street-block that have access to electricity. Such data complements the
raw spatial features, giving some description of the development status across neighborhoods.

Survey features are those available from the CUIS. These data include
features like the program recipients' ages, occupations, and needs.

We applied basic preprocessing to all features before modeling. For missing values in numeric features, we use
median imputation, while for categorical features we impute the most common class. All categorical features are converted to their dummy variable equivalent before training.

\subsection{Models}

For the underreporting problem, our baseline predicts that each household is not underreporting -- this is the majority class.
Across feature sets, we applied random forests (RF), varying the number of trees and choice of splitting criterion \cite{breiman2001random}. While other models could potentially improve performance, we preferred the generic applicability of random forests across various feature sets without the need for extensive hyperparameter optimization. This allowed for greater focus on feature engineering which we believe will result in more significant improvements.

For the imputation problem, our baseline predicts the majority class for each of the six indicators. 
We also used nearest neighbors (kNN), gradient boosting machines (GBM), and random forests (RF) -- nonlinear methods were more appropriate for drawing decision boundaries between spatial coordinates 
and identifying subsets of time indicative of certain programs \cite{devroye2013probabilistic,friedman2001greedy}. To account for scale,
we trained different models across the thirty-two states in Mexico, with Mexico City, the state with the most beneficiaries, further split into subregions.
This parallelization also accounts for heterogeneity in the regression function across regions.
\subsection{Evaluation}
\label{sec:evaluation}

For both the underreporting and imputation problems, we use nested cross-validation to estimate out-of-sample precision and recall; we further construct visualizations to compare model results with external factors. 

\subsection{Underreporting problem}
We first describe evaluation in the underreporting problem. To properly simulate the situation in which a model encounters a 
new household that is not part of the training data, we nest cross-validation folds at the household level. That is, every household
is contained entirely within a single cross-validation fold. 

As our response is binary, and since each of our classifiers 
provides a predicted probability for each class label, we are able to calculate precision-recall curves. 
For a fixed decision threshold, precision measures the number of 
flagged underreporters who were actual underreporters, while recall measures
the proportion of actual underreporters who were identified. The curves
are created by varying the threshold on a grid from 0 to 1.
This metric is appropriate, considering that in practice, following-up on flagged underreporters takes 
resources -- precision-recall curves help navigate the trade-off.

\subsubsection{User interface for exploring results}
To guide interpretation of model results, we implemented a Shiny  \cite{studio2014shiny} application to sort samples according to a user-adjustable loss-function. 
Specifically, it is useful to place a household's estimated underreporting probability in context of its distance from the poverty line
and the discrepancy between self-reported and estimated incomes. 
This is because underreporting is only problematic for households just above the poverty line, where underreporting could affect program enrollment.
Further, a difference between estimated and self-reported incomes would corroborate any suspected underreporting identified by our models. 
In our application, predicted underreporting probability and income discrepancies are plotted against each other on a scatterplot, and households far from the 
poverty line are faded into the background. Brushed points are printed in a table whose rows are sorted according to a tunable weighting of these three factors.
See Supplementary Figure \ref{fig:enrollment_app} for a screenshot.

\subsection{Imputation problem}
For evaluation in the imputation problem, we again nest cross-validation folds at the household level. We split evaluation across
states, since we expect prediction to be more difficult in some than others. Our 
response is a multidimensional binary vector, one for each poverty indicator. Instead of 
combining error across indicators, we compute precision and recall curves for indicators separately. Hence, we base evaluation on a grid of precision and recall 
curves across state and indicator combinations, see Figures \ref{fig:imputation_precision} and \ref{fig:imputation_recall}.

Finally, note that we consider imputation a purely descriptive exercise. 
Our goal is to impute poverty indicators on historical data in PUB, without necessarily attempting to forecast 
the value of indicators in the future, so do not pursue temporal validation.
\begin{figure}
\includegraphics[width=8cm]{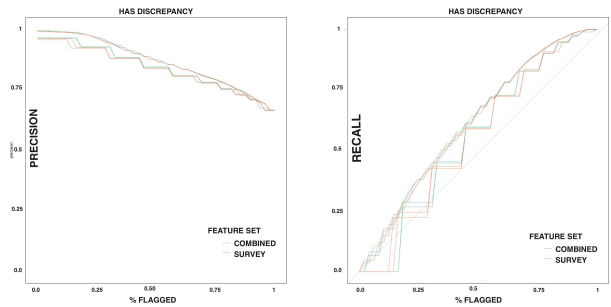}
\caption{Precision (left) and Recall curves (right) obtained from various models applied to the \texttt{any\_discrepancy} outcome in the underreporting problem using Survey and Combined (Survey + Demographic) features.}
\label{fig:underreporting_precision}
\end{figure}

\begin{figure}
\includegraphics[width=5cm]{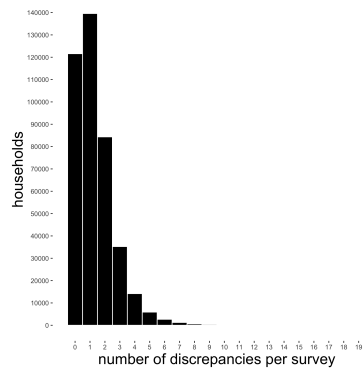}
\caption{Number of discrepancies found per survey.}
\label{fig:underreporting_num_of_discrepancies}
\end{figure}

\subsection{Results}
The results of the underreporting experiment are summarized in Figure \ref{fig:underreporting_precision} and show how different models perform using survey features and a combination of survey and demographic data.  Both feature sets use spatial data. Of the 409,000 families for whom we have home verification information, 70\% of those visits contain a discrepancy on at least one survey question which in our formulation includes both under and over reporting counts. Of those surveys with discrepancies, the majority of them, 91\%, contain 3 or fewer discrepancies, see Figure \ref{fig:underreporting_num_of_discrepancies}. This finding may be due to a respondent's misunderstanding of a question, a data collection error where the surveyor does not mark an initial question correctly, or misrepresentation by the person surveyed.  The high baseline for discrepancy likelihood is reflected by levels in the precision graph. Because of this likelihood imbalance and due to the utility value for SEDESOL, future modeling of this task will also handle individual question level responses. This task is more challenging however since most individual questions have a discrepancy level well beneath 5\%, see Supplementary Figure \ref{fig:underreporting_variables}.

\begin{figure}
\includegraphics[width=9cm]{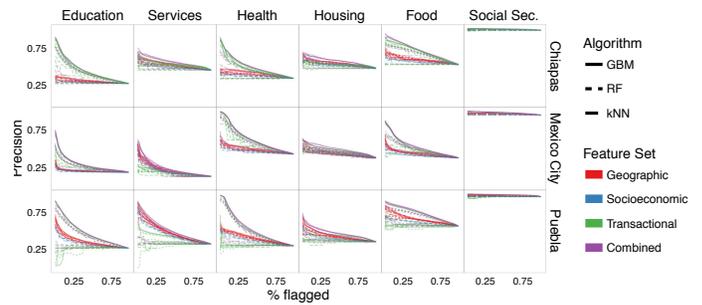}
\caption{Precision across models and feature sets, for the imputation problem. Columns give different poverty indicators, sorted from least to most prevalent, and rows correspond to different subregions (only 3 of 34 shown). Different colors give different feature sets, and line types correspond to algorithms.}
\label{fig:imputation_precision}
\end{figure}

\begin{figure}
\includegraphics[width=9cm]{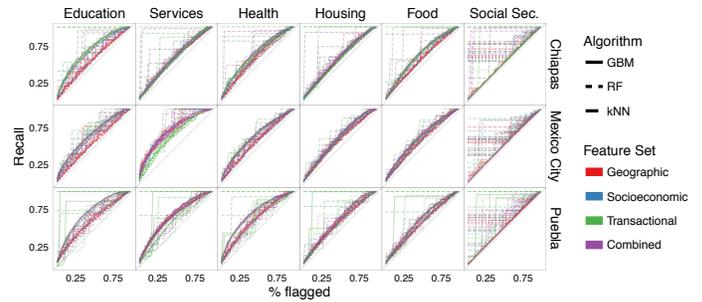}
\caption{Recall curves corresponding to the precisions in Figure \ref{fig:imputation_precision}.}
\label{fig:imputation_recall}
\end{figure}

Results for the imputation problem are summarized in Figures \ref{fig:imputation_precision} and \ref{fig:imputation_recall}. We use
kNN (12 and 25 neighbors), GBM (100 and 150 estimators), RF (50 and 100 trees) across four feature sets -- geographic, socioeconomic, transactional, and all combined -- split across 34 regions. To save space, 
only three are shown. Chiapas is an example of a rural state, while both Puebla and Mexico City\footnote{This is actually one -- the most densely populated -- of several subregions of Mexico City on which we parallelized.} are urban. As a reference, models that do not improve precision over the naive baseline, which flags all individuals as lacking an indicator, have horizontal precision curves set at the overall fraction of individuals with a given indicator. This is the exactly the 100\%-flagged value of the precision curves in Figure \ref{fig:imputation_precision}.
Similarly, the reference recall curves of a naive baseline that randomly flags individuals without regard to their features would have exactly diagonal curves in Figure \ref{fig:imputation_recall}.

We note two filtering steps used to simplify the problem. First, for some poverty indicators,
a small fraction of samples in CUIS are missing a label, we discard these in both training and evaluation.
Second, while we use the potentially noisy geocoding results based on samples with partially missing addresses, we discard those 
with no locality-level information available. This reduces the data from 18 million to 7.4 million individuals. For the remaining households -- between 30,000 and 550,000 for each state -- we 
use the nested CV approach described before, visualizing precision and recall on every held out data set.

Baselines are quite different across indicators and states, which can be read from Figure \ref{fig:imputation_precision}. For example, the rightmost column indicates that almost all beneficiaries lack access to social security, and our models provide little benefit over flagging everyone with the deprivation. Conversely, when a state has fewer
people with an indicator, our model has more value; see the difference between prediction of access to services between Mexico City, a more urban area where lack of access to utilities is rarer, and Chiapas, a state with relatively less developed infrastructure.

Different feature sets are more or less informative, depending on the indicator being analyzed. For example, program enrollment features are useful for predicting access to education, health, and food deprivations, while these same features are not  
as useful as geographic information when predicting access to services and adequate housing. In retrospect, this relationship is natural -- access to housing and services like electricity and garbage disposal would be expected to be associated with geography, while 
education, health, and food access tend to be targeted by specific programs within SEDESOL.

Among the three models, GBM and RF deliver comparable performance, and are consistently better than kNN. We note that hyperparameters have not been extensively tuned, and are chosen in order to allow parallel training of multiple models on machines with modest computational resources within an hour.

\section{Discussion}

To understand how the models generate predictions, we inspect the feature importances of successful models and visualize interclass variation for the more important features.

\begin{figure}
\includegraphics[width=6.5cm]{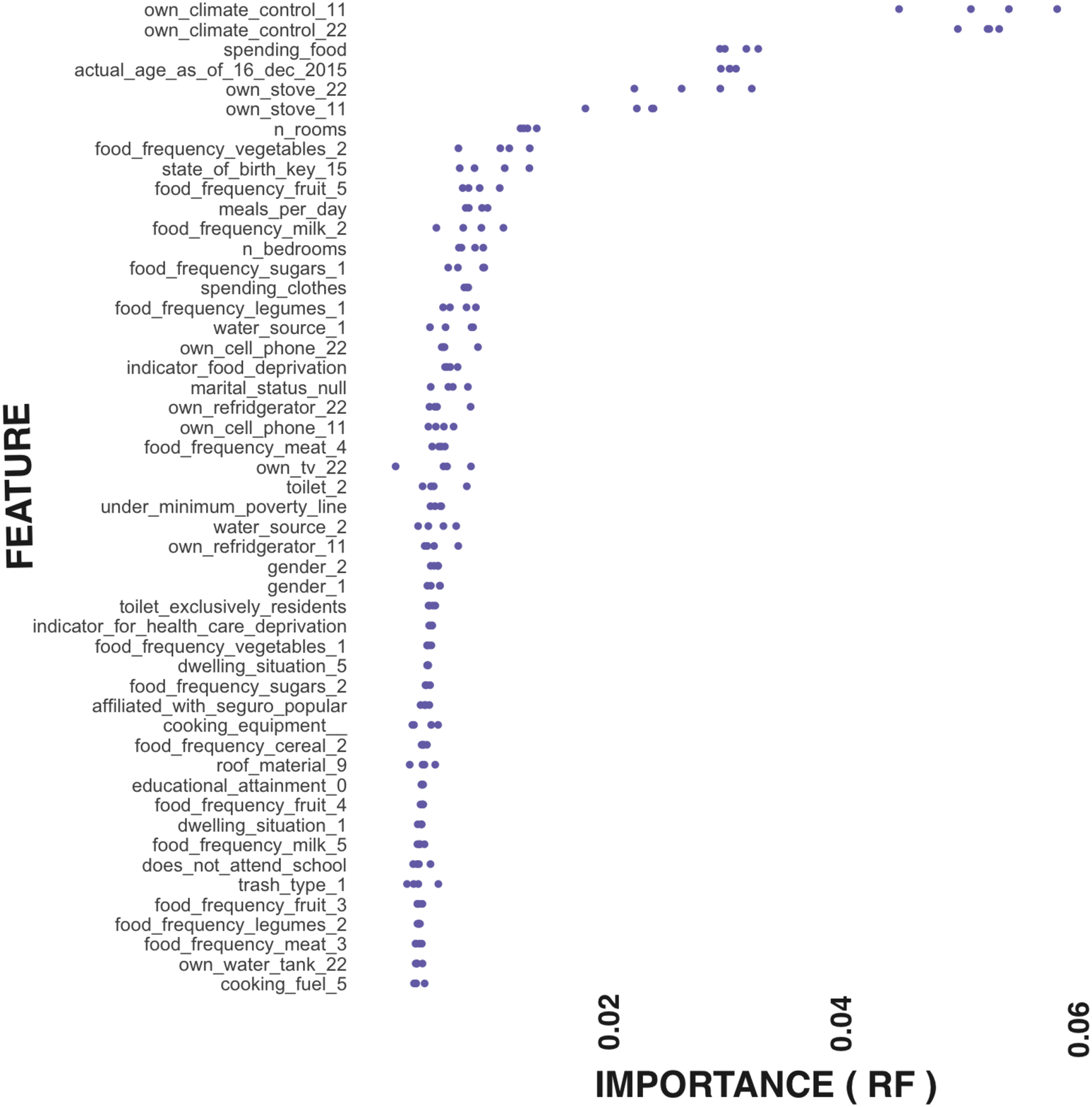}
\caption{Feature importances from the RF applied to the underreporting problem using survey, socioeconomic, and spatial data.}
\label{fig:rf_importances_underreport}
\end{figure}

Feature importances for RFs with either 10 or 50 trees using combinations of survey, socioeconomic, and spatial features associated with the underreporting problem are displayed in Figure \ref{fig:rf_importances_underreport}. Air conditioning and stove ownership, money spent on food, age of the person surveyed, number of rooms reported, the frequency of consumption of  vegetables, milk and fruit, and meals per day were the most important features.

Of the approximately 130,000 cases where there was a discrepancy with stove ownership, 127,000 were cases of overreporting, i.e. where the survey respondent said they owned a functioning stove when they did not. The true underreporting frequency for stove ownership is actually less than 2.5\%! Similarly for air conditioning ownership, only in 1000 out of 95,000 cases where a discrepancy had been found was due to underreporting. 

These ``dignity'' discrepancies, where a respondent misreports their living situation, were mentioned by surveyors during a site visit to the country.  Being born in the Federal State of Mexico is also curiously included in the group of important features, though that seems more a consequence of it being the most populous state in the country.

\begin{figure}
\includegraphics[width=6.5cm]{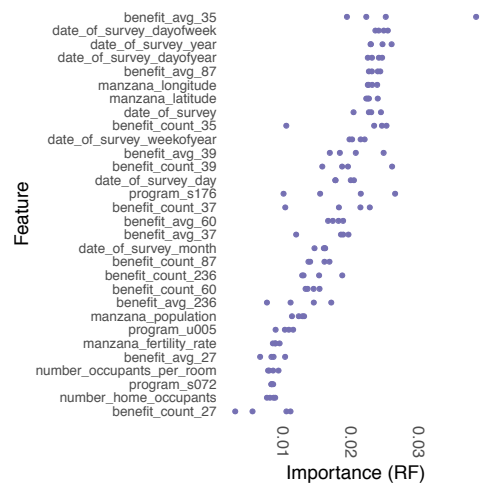}
\caption{Feature importances from the RFs, across multiple folds and parameter settings, trained on the combined feature set, when predicting access to education in Chiapas.}
\label{fig:rf_importances_imputation}
\end{figure}

\begin{figure}
\includegraphics[width=6.5cm]{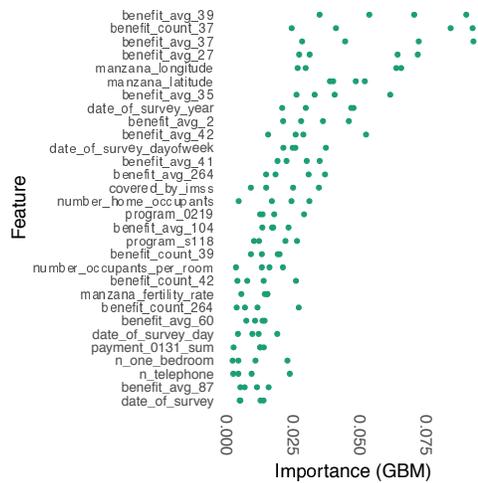}
\caption{The analog of Figure \ref{fig:rf_importances_imputation} computed from the corresponding random forest models.}
\label{fig:gbm_importances_imputation}
\end{figure}

Feature importances associated with prediction for education access in Chiapas based on the combined feature set are displayed in Figure \ref{fig:gbm_importances_imputation} and Supplementary Figure \ref{fig:rf_importances_imputation}.
Only the top 50 features are displayed. The GBMs concentrate on a smaller subset of features, compared to the RFs, but there is high overlap between the top features between
the two models. We know from Figures \ref{fig:imputation_precision} and \ref{fig:imputation_recall} that transactional features are likely most informative,
but we can now identify specific programs and benefits associated with education access. Most socioeconomic variables seem only weakly informative, suggesting limited utility of census data alone. Finally, 
even though geographic features on their own are relatively poor predictors, the households' street block coordinate (\texttt{manzana\_latitude} and \texttt{manzana\_longitude}) appears among the top ten predictors for
both GBMs and RFs.

\begin{figure}
\includegraphics[width=6cm]{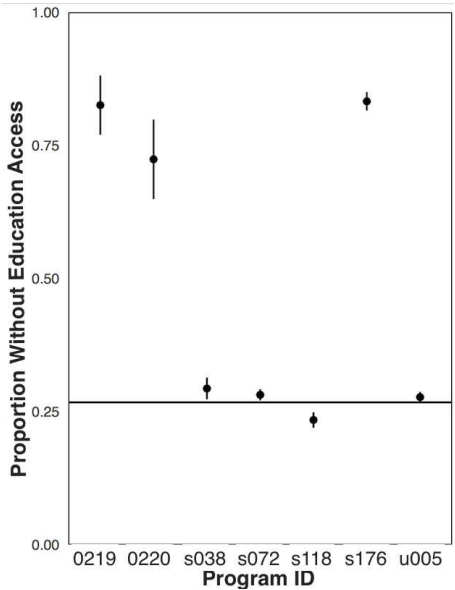}
\caption{How does program enrollment relate to education access? From transactional data, we can estimate the 
proportion of households that do and do not have access to education, grouping by program. The vertical bars give a 95\% confidence interval on estimated 
proportions -- some programs have more beneficiaries, and have more precise estimates. The horizontal line is the proportion of households with this indicator, across all samples.}
\label{fig:programs}
\end{figure}

\begin{figure}
\includegraphics[width=8cm]{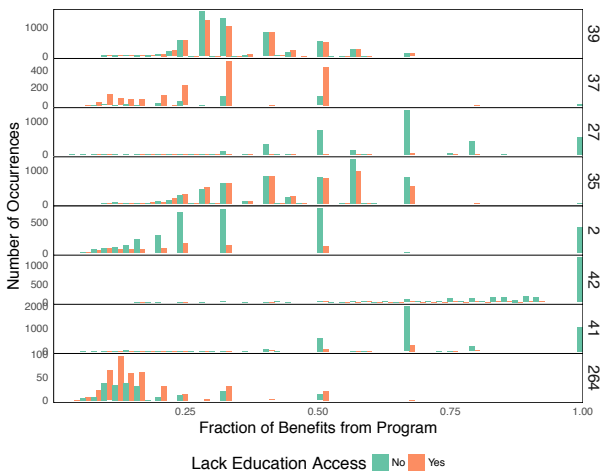}
\caption{How do program payments relate to education access? Each row is a SEDESOL program, and the 
colors in the histograms correspond to households with and 
without access education. When all of a household's benefits come from a given program, it falls into the bin at the far right for that program. Spikes at 0, for households never receiving the associated benefit, are omitted.}
\label{fig:benefits}
\end{figure}

The programs and benefits appearing in these top variables are displayed in 
Figures \ref{fig:programs} and \ref{fig:benefits}, respectively. Figure \ref{fig:programs}, 
describes programs whose beneficiaries are either more or less likely to have the education deprivation, compared to a random household. For example, program S176 
is Pensi\'on para Adultos Mayores, a pension program for senior citizens. Those who are in the program are more likely to have the education 
deprivation\footnote{Note that the access to education field asks for the highest education level of different household members. It is possible that seniors enrolled in this program never attained higher levels of education.}. The program
with a lower proportion lacking access is a food program (Programa Apoyo Alimentario), and it is possible that households in this program
are targeted for insufficiency on the food dimension of poverty, rather
than the education dimension.

Figure \ref{fig:programs} displays the frequency of benefits from programs with high variable importance, split according to whether the household is flagged as lacking access to education. Since most households do not receive 
these benefits, there is a large spike at zero for each row -- we omit this for clarity. We can see the discriminative potential 
in the way different colors are not split evenly with the red bars (lacking access) at about 1/2 the height of the green bars (not lacking access), as would be expected by chance, according to the baseline from Figure \ref{fig:programs}. 
For example, if a household received any benefits from programs 39 (a renewable energy program), 27 (Apoyo por concepto de beca, a scholarship program), or 2 (Litro de leche, a milk distribution program), etc. we would suspect that they do not lack access to education. On the other hand, if they received 
benefits from programs 37 (Apoyo para adulto mayor, a program for senior citizens), or 264 
(Servicios educativos de alfabetizacion, a literacy program), it is likely that they lack access to education in their multidimensional profile.

\begin{figure}
\includegraphics[width=8cm]{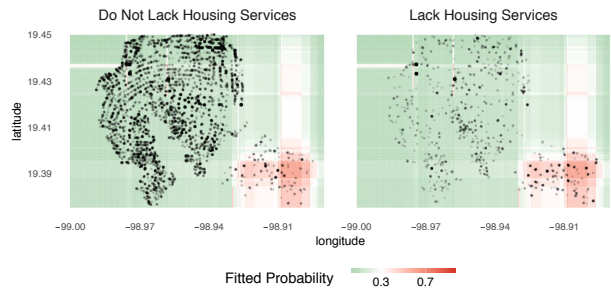}
\caption{Points here correspond to the street-block coordinates of different households in a neighborhood within Mexico City -- darker points indicate more households at that block. The two panels 
separate households with and without access to housing services. 
The background color is the predicted probability of lacking access using a GBM trained on geographic features.}
\label{fig:pub_map}
\end{figure}

We have chosen to study education in Chiapas because it was a clear example where using transactional features yielded meaningful improvements over baseline. 
For an example where geographic features outperform program features -- the main alternative regime visible in the precision and recall 
curves -- consider prediction of services in Mexico City. In Figure \ref{fig:pub_map} we plot a subset of the Mexico City model's training data, split across whether the household lacks access to services and shaded by the predicted
probability of lacking access according to a RF trained on geographic features. Since many households can map to the same street 
block, we make points semitransparent and jitter slightly. A few dark clumps are associated with 
errors in geocoding -- the center of a neighborhood is sometimes returned when no higher-resolution location can be found.
Nonetheless, it appears that the method has identified a neighborhood, in the bottom right, whose residents seem
to more frequently lack access to services.

\section{Conclusion}
We have systematically analyzed the problems of flagging underreporters and imputing poverty indicators, which has consequences for
the design and delivery of services at SEDESOL. 

We found that survey data alone can suggest potential underreporting, and have conjectured that much of what appears to be underreporting is related to misunderstanding. Further, we found geographic features useful for housing and service related indicators and transactional data informative for other dimensions of poverty.

To deploy this system in the field, SEDESOL is updating a mobile phone application, used by surveyors delivering the CUIS, so that it can send requests to an API providing historical survey data. 
Ultimately, the goal is to incorporate a version of the underreporting model presented here 
into the application, so that problematic responses could be flagged and followed-up in real time.

We have built and evaluated a machine learning system for estimating poverty profiles based on training data from 7.4 million individuals. To the best of our knowledge, this is 
the first time data with this scale and granularity has been leveraged to improve service distribution for poverty alleviation at the national level. Nonetheless, there 
are still unused data sources with the potential to improve predictive performance, for example, the 10.6 million people with CUIS and PUB records for whom no locality information is available.

We believe that this is an effective and efficient use of administrative data that social development are already collecting that can now be used to improve their operations and better achieve their goals. We hope our approach provides a further example for other agencies around the world seeking to use data to better serve those in need.

\section*{Acknowledgements}
%Blind acknowledgements for review
This work was  done as part of the 2016 Eric \& Wendy Schmidt Data Science for Social Good Summer Fellowship at the University of Chicago.

\bibliographystyle{aaai}
\bibliography{bibliography}

\clearpage

\begin{figure}
\includegraphics[width=18cm]{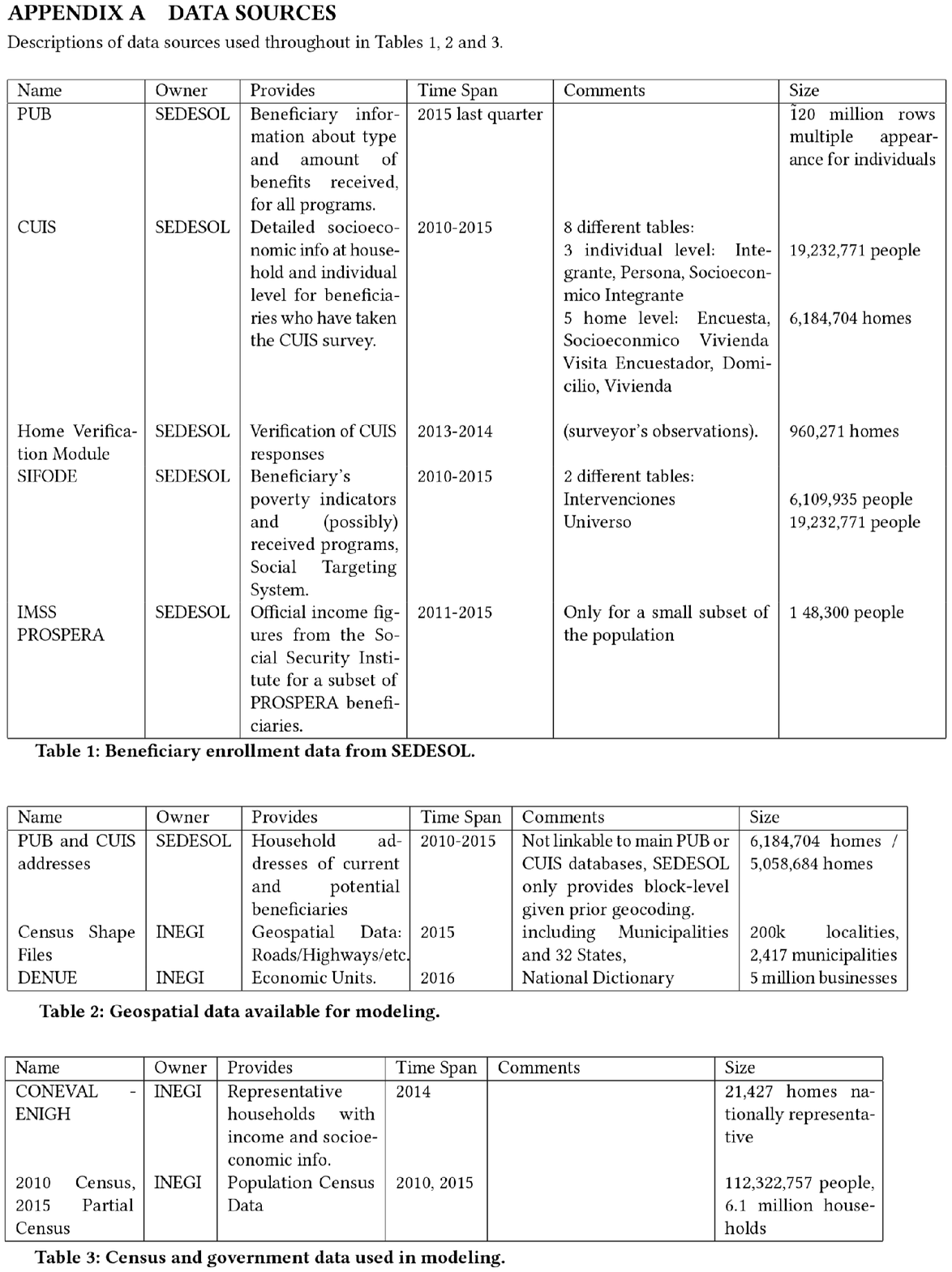}
%\caption{Data sources used}
\label{fig:data_sources}
\end{figure}
\clearpage
\begin{figure}
\includegraphics[width=6cm]{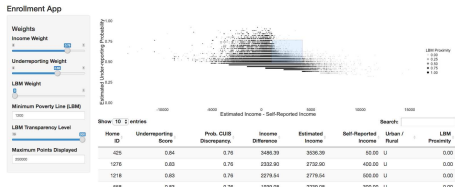}
\caption{A screenshot from the Shiny app used explore results from the underreporting model.}
\label{fig:enrollment_app}
\end{figure}

\begin{figure}
\includegraphics[width=6cm]{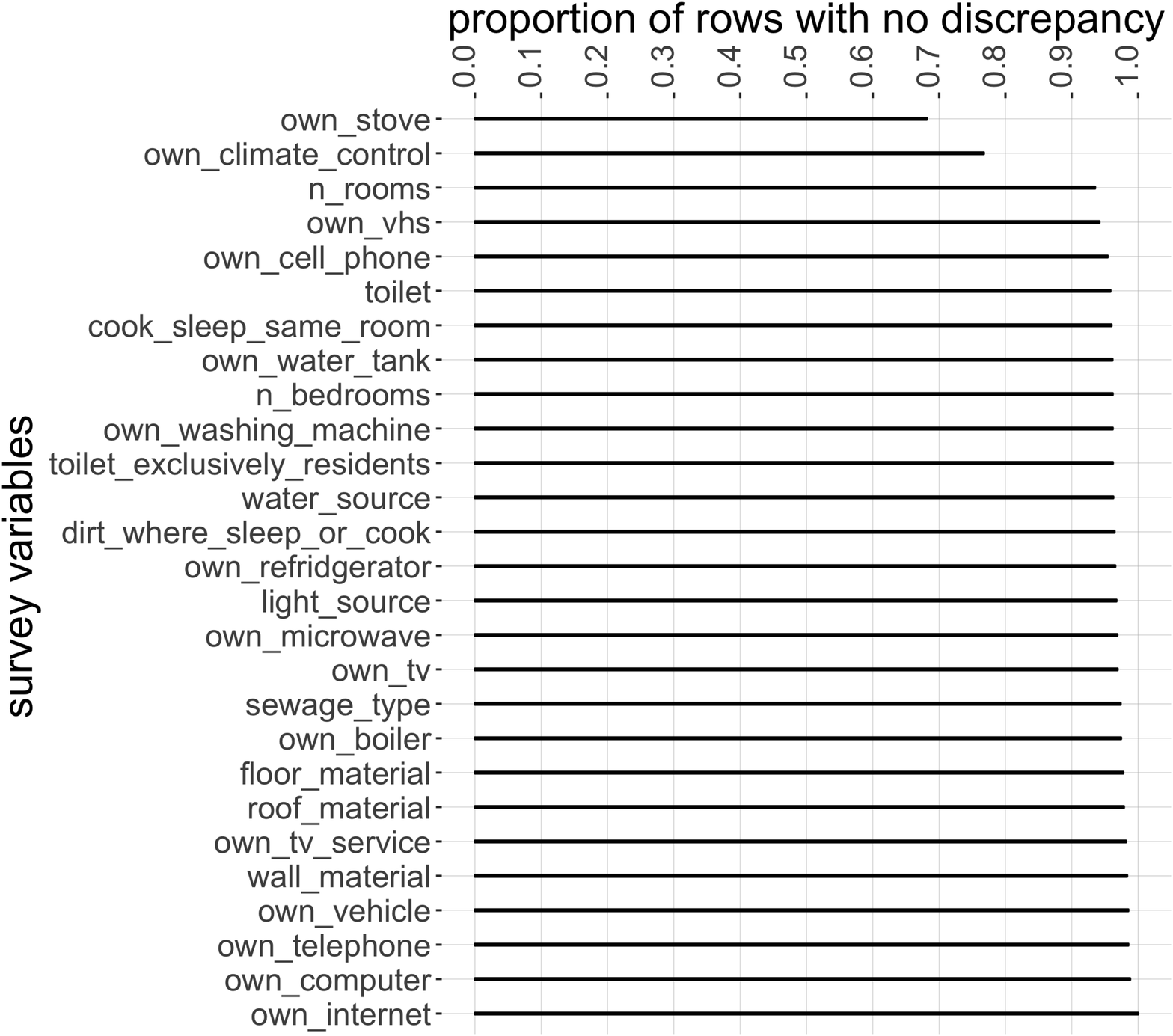}
\caption{Proportion of survey questions without discrepancies.}
\label{fig:underreporting_variables}
\end{figure}

\end{document}